\documentclass[a4paper,fleqn,usenatbib]{mnras}
\usepackage{amsmath}
\usepackage{epsfig}
\usepackage{graphicx}
\usepackage{capt-of}
\usepackage{color}

\usepackage[T1]{fontenc}
\usepackage{ae,aecompl}

\begin{document}

\title{Envelopes for orbits around axially symmetric sources with spheroidal shape}

\author[J. Ramos-Caro and R. S. S. Vieira]
{Javier Ramos-Caro$^{1}$\thanks{E-mail: javier@ufscar.br} and Ronaldo S. S. Vieira$^{2}$\thanks{E-mail: ronaldo.vieira@ufabc.edu.br}\\
$^{1}$Departamento de F\'{i}sica, Universidade
Federal de S\~{a}o Carlos, 13565-905, SP, Brazil\\
$^{2}$Centro de Ci\^encias Naturais e Humanas, Universidade Federal do ABC, 09210-580 Santo Andr\'e, SP, Brazil}

\date{Accepted XXX. Received YYY; in original form ZZZ}

\pubyear{2023}

\maketitle

\begin{abstract}

We introduce a method to obtain the envelopes of eccentric orbits in axially symmetric potentials, $\Phi(R,z)$,
 endowed with $z$-symmetry of reflection. By making the transformation $z\rightarrow a+\sqrt{a^{2}+ z^{2}}$,
 with $a>0$, we compute the resulting mass density, referred here as the \emph{effective density} $\rho_{\rm ef}(R,z;a)$, in order to
  calculate the envelopes $Z(R)$ of orbits in the meridional plane $(R,z)$. We find that they  obey the
  approximated formula $Z(R)\propto [\Sigma_{\rm ef}(R;a\approx 0)]^{-1/3}$, where $\Sigma_{\rm ef}(R;a)$ is the integrated surface density
  associated with $\rho_{\rm ef}(R,z;a)$.
As examples we consider the dynamics in two potentials: the monopole plus quadrupole and the Kalnajs disc.

\end{abstract}


\section{Introduction} \label{sec:intro}

One of the paradigmatic problems of the theory of orbits is the existence of a non-classical third integral of motion
in axisymmetric potentials.
Numerical experiments have demonstrated that the vast majority of situations are characterized
by a phase space with prominent regions of regular orbits, even beyond the predictions
of linear approximations and the Kolmogorov-Arnold-Moser (KAM) theory. Indeed, we can find (approximate) integrability even in
situations where these approaches cannot be applied, as for example, in the case of
disc-crossing orbits in galactic models involving razor-thin discs,
due to the discontinuity of the gravitational field on the equatorial plane. Although it is not possible to apply the epicyclic approximation and
current versions of KAM theorem,
the corresponding Poincar\'{e} surfaces
 of section have frequently shown large areas of regularity
 (see for example \citealp{saa,hunter,ramos-caro2,pedraza}).

This issue has motivated several authors to undertake the search for an explicit expression
for the third integral of motion (e.g., \citealp{Contopoulos2,Contopoulos1}).
In some cases this search has been fruitful but there are a lot of situations in which this task
has became very difficult \citep{boccaletti}. However, it is always possible to work on the basis of approximative
schemes (in regions filled by regular orbits) which can facilitate this search. One important example is the so-called adiabatic approximation, very useful
to perform accurate estimations for  orbital envelopes in galactic models \citep{bt,binney,sanders}, an important
ingredient in the formulation of dynamical models for the Galaxy \citep{binney}.

Another interesting example is manifested in the context of  disc-crossing orbits, where the authors
found a way to deal with the discontinuity in the gravitational field
of such models, providing an alternative scheme to perform the approximations, which leads to a simple expression for the
third integral associated to disc-crossing orbits of small z-direction (vertical) amplitude, given in cylindrical coordinates ($R, \varphi,z$) by
\begin{equation}
I=\frac{1}{2\pi G} \left[\Sigma(R)\right]^{-2/3}\,\left[\frac{1}{2} P_z^2 + 2\pi G\,\Sigma(R)\,|z|\right]\,,
\end{equation}
where $\Sigma(R)$ is the surface mass density of the razor-thin disc \citep{vieira}. When evaluated at the orbit's envelope $Z(R) = z_{max}(R)$, we have  $I=Z\Sigma^{1/3}$, in such a way that $Z(R)\propto\left[\Sigma(R)\right]^{-1/3}$.
Numerical experiments with the Kuzmin
and Kalnajs discs showed that the above expression for the envelopes is also valid for orbits with large vertical amplitude
(i.e. of the  order of radial amplitude). A similar statement holds for
extended models including also a three-dimensional (3D) disc and other components,
but instead of a surface density $\Sigma$, the envelopes are
proportional  to the
integrated dynamical surface mass density $\Sigma_{I}$ \citep{vieira2}, given by
\begin{equation}\label{DensidadeIntegrada}
\Sigma_{I}(R)=\int_{-\zeta}^{\zeta}\rho_{\mbox{tot}}(R,z)dz,
\end{equation}
where $\rho_{\mbox{tot}}$ is the total density of the $3$D-configuration (i.e. the sum of thin disc, thick disc, bulge and spheroidal halo densities)
 and $\zeta$ is the thickness of the thin disc, which can be considered as a constant for sufficiently flattened discs. We then found that envelopes
 of box orbits with amplitudes of the order of thick disc are accurately predicted by a formula similar to that used for thin discs:
 $Z\propto \Sigma_I^{-1/3}$.
Subsequently, such approach was extended to describe the envelopes of regular orbits with arbitrary vertical amplitudes inside the disc distribution
\citep{vieira3}, improving the predictions of the adiabatic approximation.

The idea that vertical amplitudes of disc-crossing orbits are inversely proportional to the cube root of an integrated density
can be extended even to situations where there is no disc, that is, for orbits in regions free of matter or in vacuum solutions
(i.e. axially symmetric solutions of the Laplace equation). But how can this be possible, taking into account that the existence  of a disc
is necessary to define an integrated density?
This paper is intended to illustrate this apparently
paradoxical fact. Since we shall deal with orbits that take place in vacuum,
we have to introduce an \emph{effective integrated density}, $\Sigma_{\rm ef}$, obtained from the vacuum solution, after being subjected to a certain
transformation (section \ref{sec:integral}). The principal effect of such a transformation is to ``create'' the 3D disc that allows us
to define such $\Sigma_{\rm ef}$ that is necessary to model the envelopes of eccentric orbits crossing the equatorial plane. As we shall see,
the predictions of this procedure (which can be summarized in the formulae (\ref{thirdIntegral})--(\ref{Miyamoto-trans2})) are highly accurate and
substantially improve  the results of adiabatic approximation.

The paper is organized as follows. In Section~\ref{sec:eqsmotion} we present the general form of the gravitational potentials we are going to focus,
along with the corresponding equations of motion for test particles. Then we introduce, in Section~~\ref{sec:integral},
the procedure to find the expression that helps us to fit the envelopes of box orbits crossing the equatorial plane.
Section~\ref{sec:comparison} is devoted to the comparison of our results with the predictions of the classical adiabatic approximation.
In Section~\ref{sec:conclusion}
we summarize the main results obtained and outline several perspectives that can be developed from them, as well as possible applications.
Finally, a word about notation: Since here we are focusing on axially symmetric potentials,
we will use cylindrical coordinates $(R,\varphi,z)$ throughout the text. Also,
we will use the
traditional notation of upper dots representing differentiation with respect to time (for example, $\dot{z}=dz/dt$ and $\ddot{R}=d^{2}R/dt^{2}$).

\section{Equations of motion around spheroidal distributions with axial symmetry}\label{sec:eqsmotion}

The external gravitational potential of a finite mass distribution endowed with axial symmetry (i.e. independent of the  coordinate $\varphi$)
can be written as (see for example \citealp{boccaletti}, pp 350)
\begin{equation}\label{1}
    \Phi=-\frac{GM}{r}\left[1+\sum_{k=2}^{\infty}A_{k}\left(\frac{R_{o}}{r}\right)^{k}
    P_{k}\left(\cos\theta\right)\right]
\end{equation}
where $M$ is the total mass of the central body, $R_{o}$ is its equatorial radius, $P_{k}$ are the Legendre polynomials and
$A_{k}$ are the coefficients of the expansion.
The above expression is given in spherical coordinates $(r,\theta,\varphi)$, which are related with the cylindrical ones by
$$
r=\sqrt{z^{2}+R^{2}},\qquad \cos\theta= z/\sqrt{z^{2}+R^{2}}
$$

If the distribution also has spheroidal symmetry, the
corresponding potential is symmetric about the plane $z=0$, i.e.
$$
\Phi(R,z)=\Phi(R,-z),
$$
which can also be referred as a $z$-symmetry of reflection. In such a case, the odd terms vanish in expansion (\ref{1})
and the potential can be written as
\begin{equation}\label{2}
    \Phi=-\frac{GM}{\sqrt{z^{2}+R^{2}}}\left[1+\sum_{n=1}^{\infty}A_{2n}R_{o}^{2n}
    \frac{P_{2n}\left(z/\sqrt{z^{2}+R^{2}}\right)}{\left(z^{2}+R^{2}\right)^{n}}
    \right]
\end{equation}
Since this potential is axially symmetric, orbits of test particles
 have two first integrals of motion, the energy $E$ and the
$z$-component of the angular momentum, $\ell$ (here we work with the specific values of these quantities, i.e. divided
by the mass of the test particle) and are determined by the Hamiltonian
\begin{equation}
   H = \frac{P_R^2 + P_z^2}{2}  + \Phi_{ef}(R, z), \label{hamiltonian}
\end{equation}
where $P_R=\dot{R}$, $P_z=\dot{z}$  and $\Phi_{ef}$ is
the effective potential, defined by
  \begin{equation}
   \Phi_{ef}(R, z) \equiv \Phi(R, z) + \frac{\ell^2}{2R^2}.
  \end{equation}
The resulting equations of motion can be written in terms of this effective potential:
  \begin{equation}\label{Req}
   \ddot{R} = -\frac{\partial \Phi_{ef}}{\partial R},
  \qquad
   \ddot{z} = - \frac{\partial \Phi_{ef}}{\partial z}.
  \end{equation}
As it was mentioned in the introduction, numerical integration of the above equations suggests, for a variety
of systems and of initial conditions, the existence of a (non-classical) third integral of motion, besides $\ell$ and $E$ (the Hamiltonian of equation~(\ref{hamiltonian})), at least in a substantial region of phase space around circular orbits.
In the next section we will deal with the problem of formulating an approximated expression for
the corresponding orbits' envelopes $Z(R)$ in the context of effective surface densities.

\section{Envelopes for Bounded Orbits}\label{sec:integral}

Since the potential of equation (\ref{2}) represents a solution of  Laplace's equation (a vacuum solution), an
appropriate method to describe orbital envelopes is the scheme of adiabatic invariants (see for example \citealp{bt}), which leads to the
following approximated expression for the envelopes of disc-crossing orbits:
\begin{equation}
Z(R)\propto[\Phi_{zz}(R,0)]^{-1/4}. \label{adiabatic}
\end{equation}
Here $\Phi_{zz}\equiv\partial^{2}\Phi/\partial z^{2}$ and $Z(R)$ represents the orbit's vertical amplitude at a radial distance $R$.
We will refer to equation~(\ref{adiabatic}) as the \emph{adiabatic approximation}.
However, it can be shown that such expression does not work well for orbits with large vertical
amplitude, as for instance, is presented in \citet{vieira2}. Indeed, as previously mentioned in the introduction,
 one remarkable result of such reference is to show that the formula $Z\propto \Sigma_I^{-1/3}$ improves the approximation of
(\ref{adiabatic}), especially when we are dealing with situations where it is possible to define an integrated dynamical surface mass density
 $\Sigma_{I}$, given by (\ref{DensidadeIntegrada}). But for the case of orbits described entirely in vacuum one cannot define a $\Sigma_{I}$, so
 the scheme presented in  \citet{vieira2} is not applicable.

However, one way to circumvent this problem is by carrying out a transformation that converts the vacuum solutions into solutions with matter,
characterized by an \emph{effective integrated density}, $\Sigma_{\rm ef}$.
If the transformation is defined in such a way that the new matter solution deviates slightly
from the original vacuum  solution, then, presumably, the corresponding orbits  will also be approximately  equivalent.
The advantage is that now, with
the orbits of a matter solution we can associate an expression for $Z(R)$, proportional to $\Sigma_{\rm ef}^{1/3}$, as a description of the orbits' envelopes in the vacuum potential.

A simple realization of the procedure outlined above can be achieved through
 the so-called \emph{displace, cut, and reflect method} \citep{kuzmin, toomre},
which starts from an axially symmetric vacuum solution, $\Phi_{\mbox{v}} (R,z)$, and, by performing the transformation
\begin{equation}\label{displ-cut-reflect}
   z\rightarrow a+|z|\qquad (a \:\:\:\mbox{is a real constant}),
\end{equation}
leads to a potential generated by a disc-like source, $\Phi_{D}(R,z)=\Phi_{\mbox{v}} (R,a+|z|)$.
For the special case in which
 \begin{equation}\label{Kuzmin-likePotential}
    \Phi_{D}=\Phi_{D}(\xi), \qquad \xi=\sqrt{R^{2}+(a+|z|)^{2}},
 \end{equation}
the solution is called a Kuzmin-like potential \citep{Tohline}.
The mass density associated with such a potential, obtained via Poisson's equation,
can be written as \citep{hunter}
\begin{equation}\label{density-Kuzminlike}
    \rho=\frac{1}{4\pi G}
    \left[\frac{d^{2}\Phi_{D}}{d\xi^{2}}+\frac{2\left(1+a\,\delta(z)\right)}{\xi}\frac{d\Phi_{D}}{d\xi}\right],
\end{equation}
from which one can identify a superposition of a surface mass distribution, corresponding to the $\delta(z)$-term
(a razor-thin disc in the equatorial plane), and a volumetric mass density (a 3D disc).

When transformation (\ref{displ-cut-reflect}) is applied to the vacuum solution (\ref{2}),
we do not obtain exactly a Kuzmin-like potential, but another class of solution of Poisson's equation associated
with some density $\rho$, in a similar fashion as in equation~(\ref{density-Kuzminlike}), with a razor-thin disc component.
Such a solution could be softened, in order to facilitate the implementation of future numerical calculations,
by considering the more general transformation
\begin{equation}\label{Miyamoto-trans}
   z\rightarrow a+\sqrt{b^{2}+z^{2}},\qquad (a,b \:\:\:\mbox{real constants}),
\end{equation}
which reduces to the Kuzmin transformation of equation~(\ref{displ-cut-reflect}) for $b=0$.
Note that the Miyamoto-Nagai potential \citep{bt} is obtained by applying (\ref{Miyamoto-trans}) to the monopole solution
(i.e. by choosing $A_{2n}=0$ for all $n$ in in the sum of equation~(\ref{2})). In our case of interest, the application of (\ref{Miyamoto-trans}) on
solution (\ref{1})
leads to the gravitational potential generated by the distribution
\begin{equation}\label{effective-density}
    \rho_{\rm ef}(R,z;a,b)=\frac{1}{4\pi G}\nabla^{2}\Phi(R, a+\sqrt{b^{2}+z^{2}}).
\end{equation}
It is clear that such solution reduces to (\ref{2}) when one chooses $a=b=0$, since the $z$-dependence of the
potential (\ref{2}) is only through $z^{2}$. Then one could expect that by choosing very small values for $a$ and $b$,
the orbits associated with the resulting  potential will deviate only slightly from the orbits (with same initial conditions)
 corresponding to (\ref{1}). But, with the orbits of the potential $\Phi(R, a+\sqrt{b^{2}+z^{2}})$ we can associate
the distribution of matter (\ref{effective-density})  and, subsequently, define an effective integrated density,
\begin{equation}\label{DensIntEff}
\Sigma_{\rm ef}(R;a,b)=\int_{-b}^{b}\rho_{\rm ef}(R,z;a,b)dz,
\end{equation}
intended to determine, in a similar fashion as in  \citet{vieira2},
an approximated formula for the envelopes of orbits crossing the equatorial plane:
\begin{equation}\label{3}
   Z(R)\propto\Sigma^{-1/3}_{\rm ef}(R;a,b).
\end{equation}
Remember that $Z(R)$ represents here the vertical amplitude of the orbit when the radial coordinate is $R$,  so,
for small values of $a$ and $b$ it could provide a reasonable prediction for true envelopes in the meridional plane.
In order to verify the validity of these statements, we performed some numerical experiments, whose procedure and results are
shown in the next section.

\section{Comparison with the adiabatic approximation}
\label{sec:comparison}

Let us consider a simple, though not trivial, particular case of  equation~(\ref{2}), which consists in retaining only the first two
terms in the expansion. It represents the gravitational potential generated by the superposition of a monopolar and a quadrupolar term,
\begin{equation}\label{eq:PotQuadrupole}
    \Phi=-\frac{\alpha}{\sqrt{z^{2}+R^{2}}}-\frac{\beta(2z^{2}-R^{2})}{2(z^{2}+R^{2})^{5/2}},
\end{equation}
where we have defined $\alpha=GM$, which usually is called monopole moment, and $\beta=GM A_2 R_{o}^{2}$, the quadrupole moment,
 representing the major deviation from the spherical symmetry.
When $\beta>0$ the body has prolate deformation and for $\beta<0$ it has oblate deformation.

It is well known that test-particle motion in the potential (\ref{eq:PotQuadrupole})
can be chaotic or regular (see for example \citealp{letelier2}). Here, to check the prediction of formula (\ref{3}), we will focus only on regular
bounded orbits that can be framed inside well-defined envelopes.

The corresponding numerical calculations can be simplified by defining
some dimensionless quantities. At first, we
  define dimensionless coordinates ${\cal R}$ and ${\cal Z}$ as
  \begin{equation}\label{Rc}
   {\cal R}=\frac{R}{R_o}, \qquad {\cal Z}= \frac{z}{R_o},
\end{equation}
where $R_o$ is the equatorial radius of the source, previously introduced in (\ref{1}). Note that, since we are focusing on exterior
orbits, it is necessary to impose the condition that ${\cal R}>1$. Also we introduce
 the dimensionless effective potential and $z$-angular momentum,
 \begin{equation}\label{phidimensionless}
   \phi_{ef}=\frac{R_o}{\alpha}\Phi_{ef}, \qquad {\cal L}_z =\frac{\ell}{\sqrt{\alpha R_o}},
\end{equation}
respectively, so that we can write
\begin{equation}\label{phidimensionless2}
   \phi_{ef}= -\frac{1}{\sqrt{{\cal Z}^{2}+{\cal R}^{2}}}-\frac{p(2{\cal Z}^{2}-{\cal R}^{2})}{2({\cal Z}^{2}+{\cal R}^{2})^{5/2}}
   +\frac{{\cal L}_{z}^{2}}{2{\cal R}^{2}},
\end{equation}
where
 \begin{equation}\label{p}
   p=\frac{\beta}{\alpha R_{o}^{2}}=A_2 .
\end{equation}

\begin{figure*}
$$
\begin{array}{cc}
  (a) & (b)\\
  \epsfig{width=6cm,file=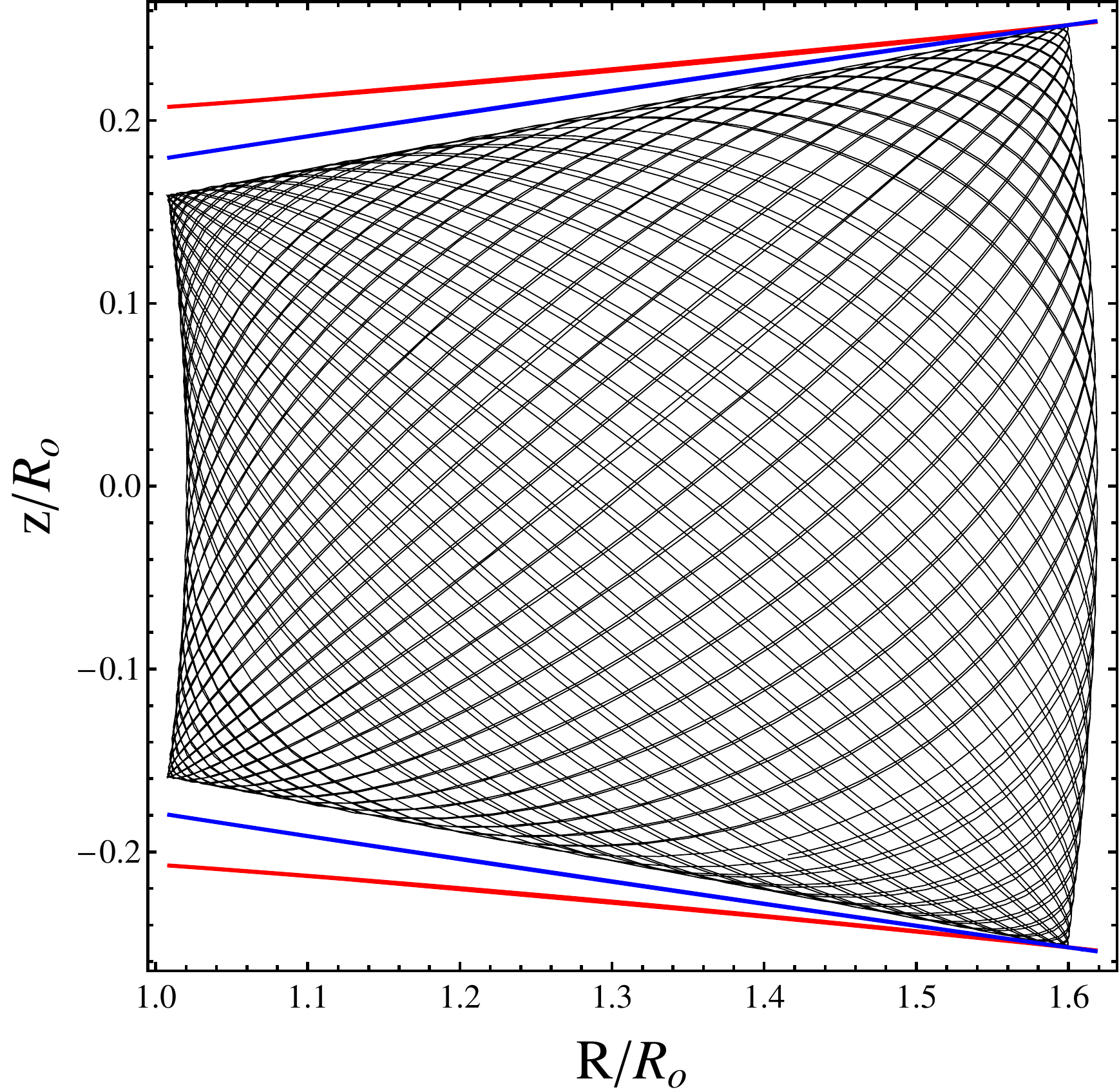} & \epsfig{width=6cm,file=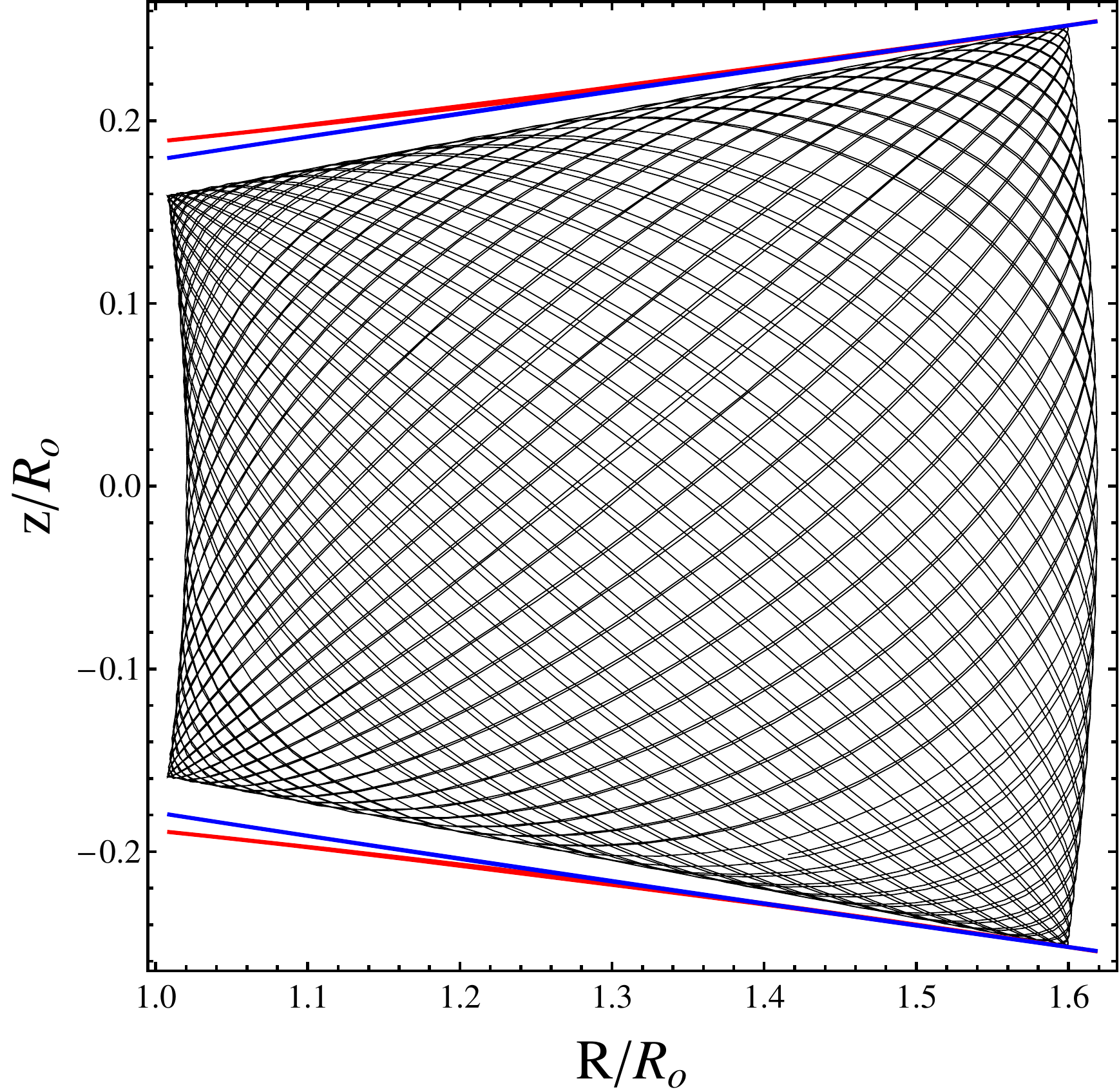} \\ \hfill \\
  (c) & (d)\\
   \epsfig{width=6cm,file=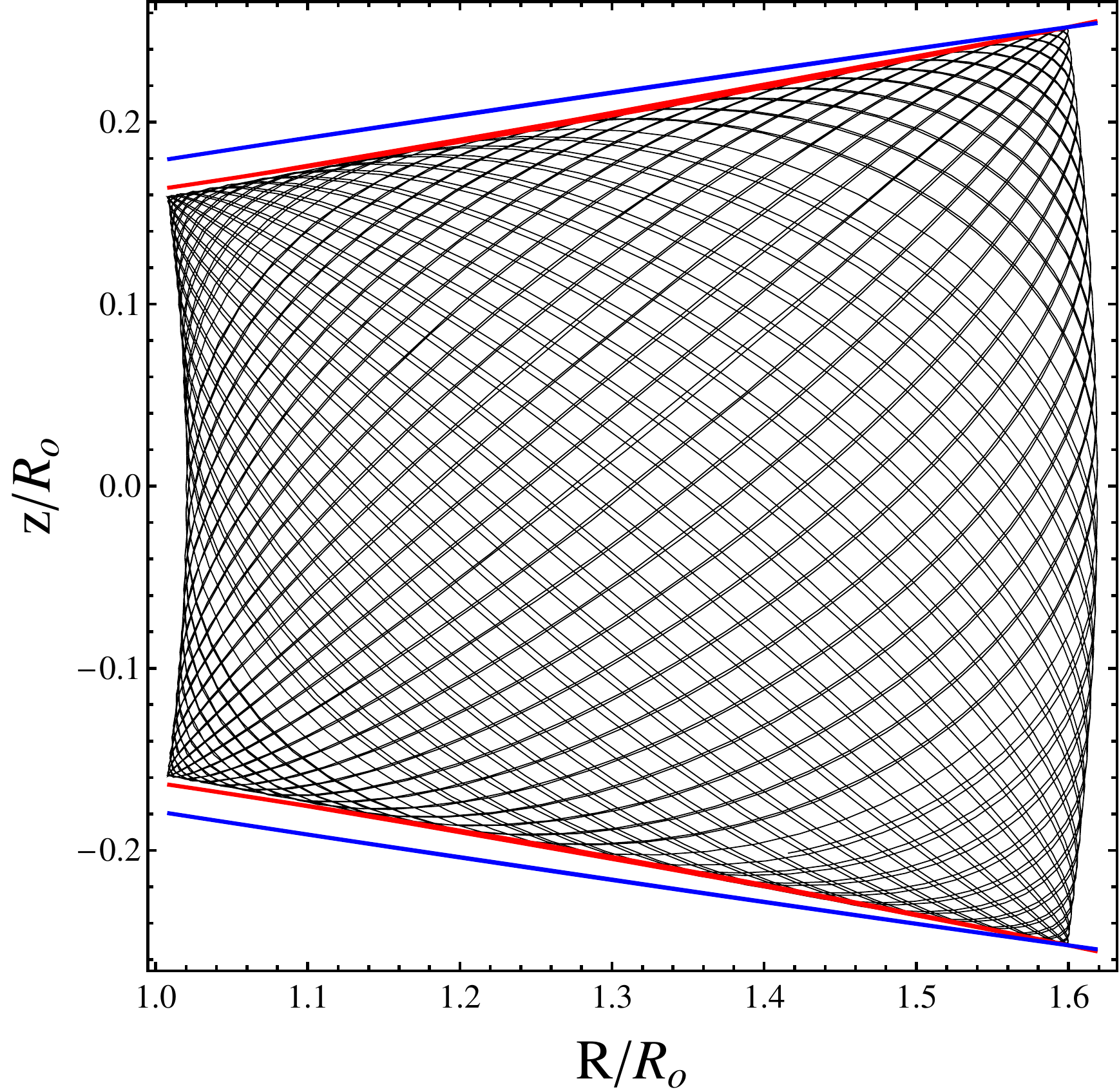} & \epsfig{width=6cm,file=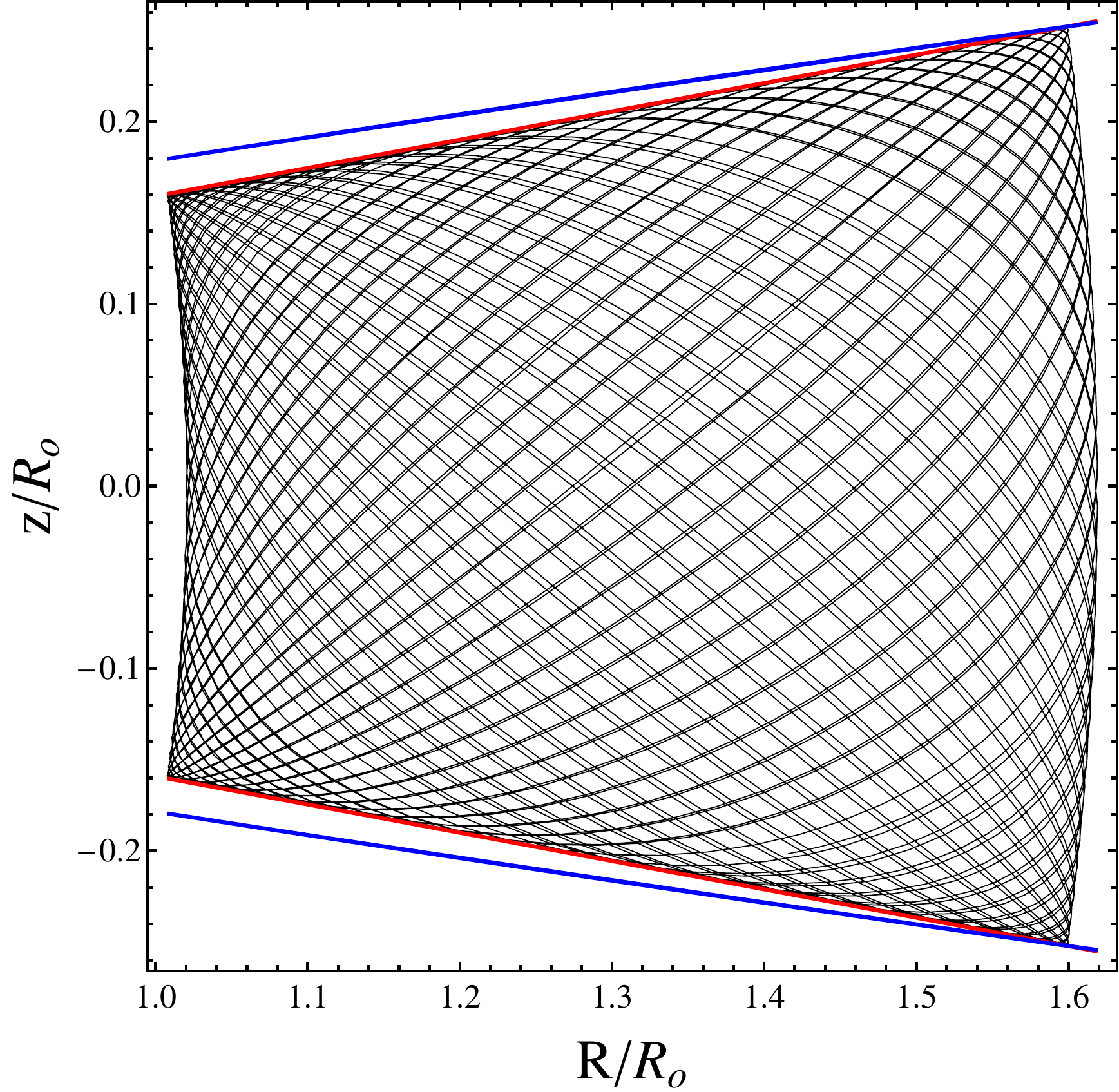}
\end{array}
$$
\caption{ Envelopes for an orbit  around a monopole plus quadrupole central body with $p=0.01$ and choosing
 $\ell/\sqrt{\alpha R_{o}}=1.1$, $E/\alpha=-0.38$, $R(0)/R_{o}=1.02$, $z(0)/R_{o}=10^{-10}$, $V_R (0)=0$.
 The numerically integrated orbit in the meridional plane $Rz$ is presented in black; the
 blue curve is the prediction of adiabatic approximation (\ref{adiabatic})
 and the red curve represents the envelope predicted by (\ref{3})
 when we choose $a = 0.01$ and (a)  $b = 2.0$, (b)$b = 1.5$, (c) $b = 1.0$, (d) $b=0.01$. The red and blue curves were plotted so that both coincide with the numerically calculated orbit as evaluated at its maximum value for $Z(R)>0$, matching precisely the zero-velocity curve at that point (top right corner of the orbit in each panel). } \label{fig:1}
\end{figure*}

In order to show the effect of parameters $a$ and $b$, appearing in (\ref{effective-density}), (\ref{DensIntEff}) and (\ref{3}),
we considered box orbits with vertical amplitudes comparable to the radial amplitudes, for which it is expected that the adiabatic
approximation (\ref{adiabatic}) does not work. At first, by choosing a small fixed value for $a$, we varied the values of $b$,
we obtain fairly generic results like those illustrated in figure \ref{fig:1}, corresponding to a typical orbit in the meridional plane $Rz$ with
$Z/R_{o}\sim 0.2 $ and envelopes with $a=0.01$. For large values of $b$ (figures \ref{fig:1}(a) and \ref{fig:1}(b)), the adiabatic
approximation (blue curve envelope), although imprecise, works better than the approximation of formula (\ref{3}) (red curve envelope).
However, the opposite happens when we select small values for $b$ (figures \ref{fig:1}(c) and \ref{fig:1}(d)), where
formula (\ref{3}) substantially improves the prediction for the orbit's envelope.

\begin{figure*}
$$
\begin{array}{cc}
  (a) & (b)\\
  \epsfig{width=6cm,file=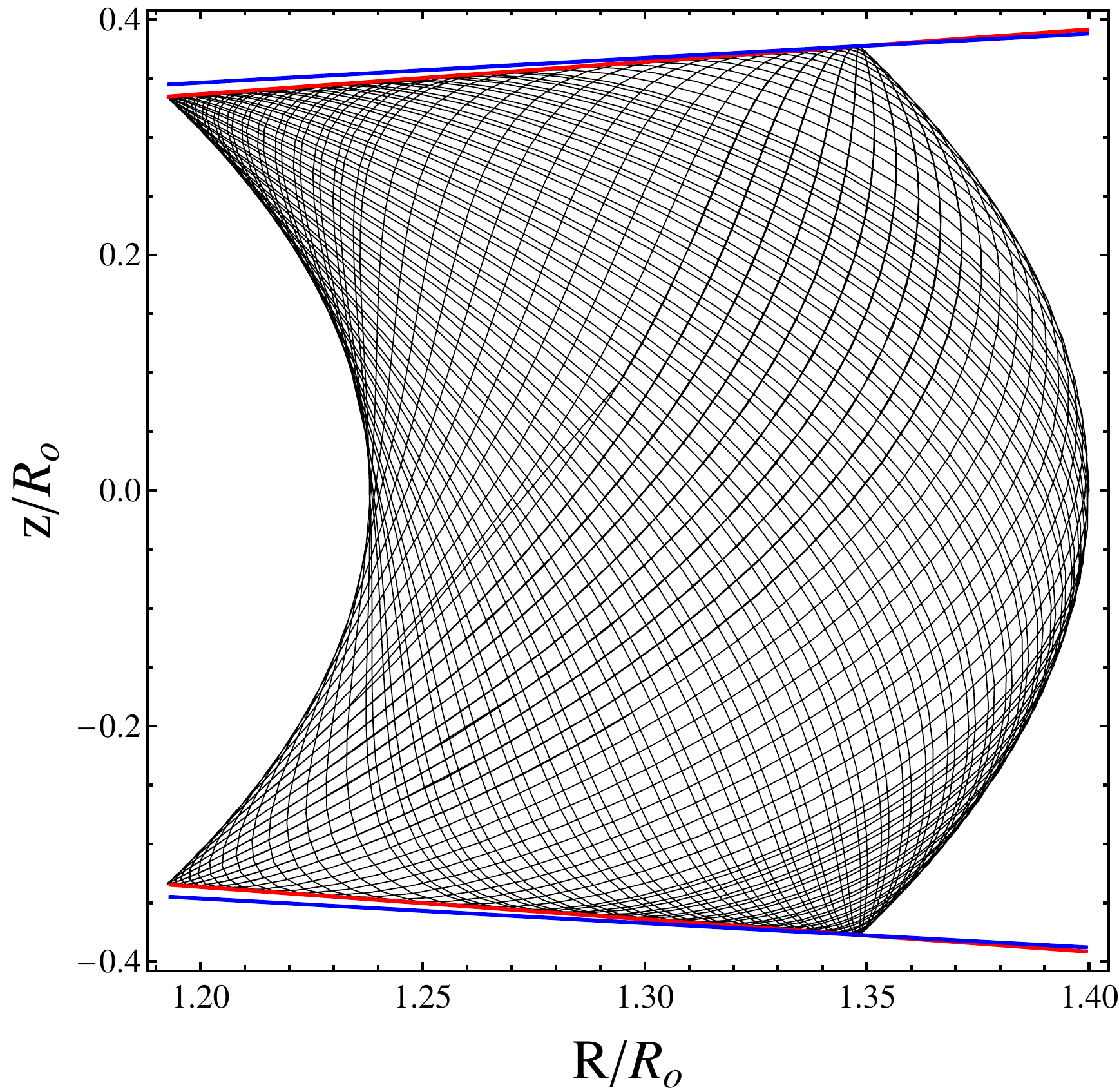} & \epsfig{width=6cm,file=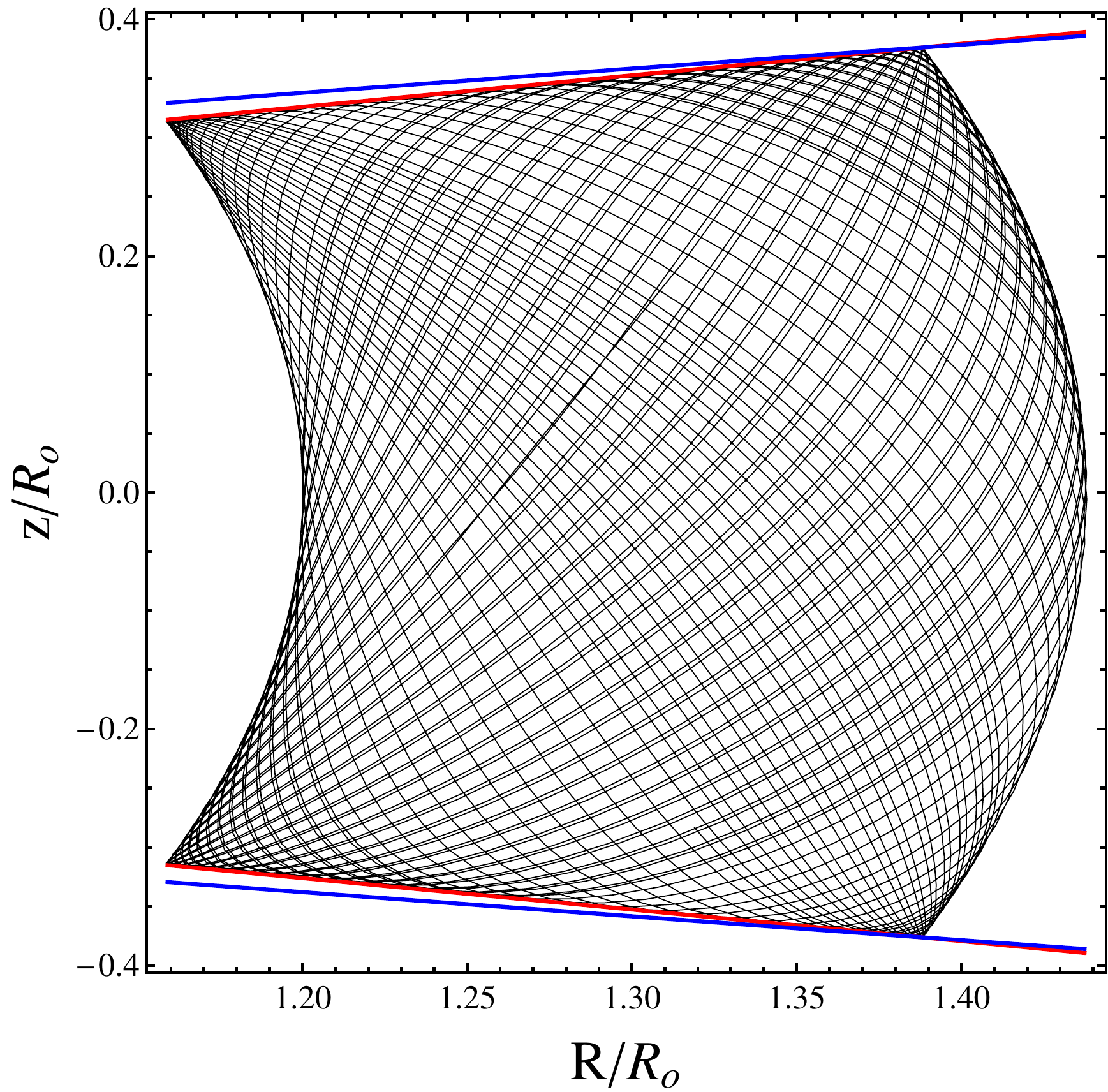} \\ \hfill \\
  (c) & (d)\\
   \epsfig{width=6cm,file=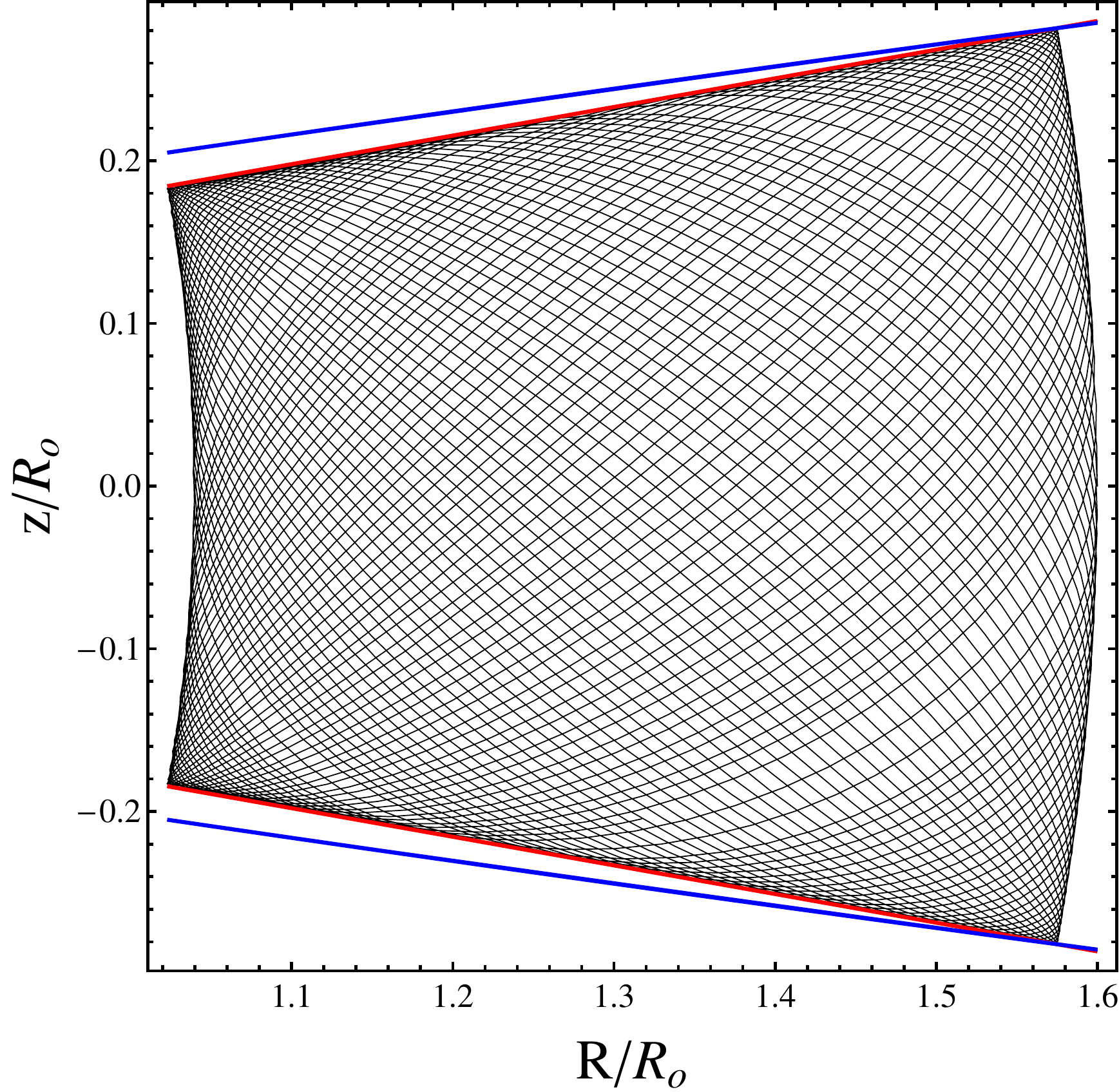} & \epsfig{width=6cm,file=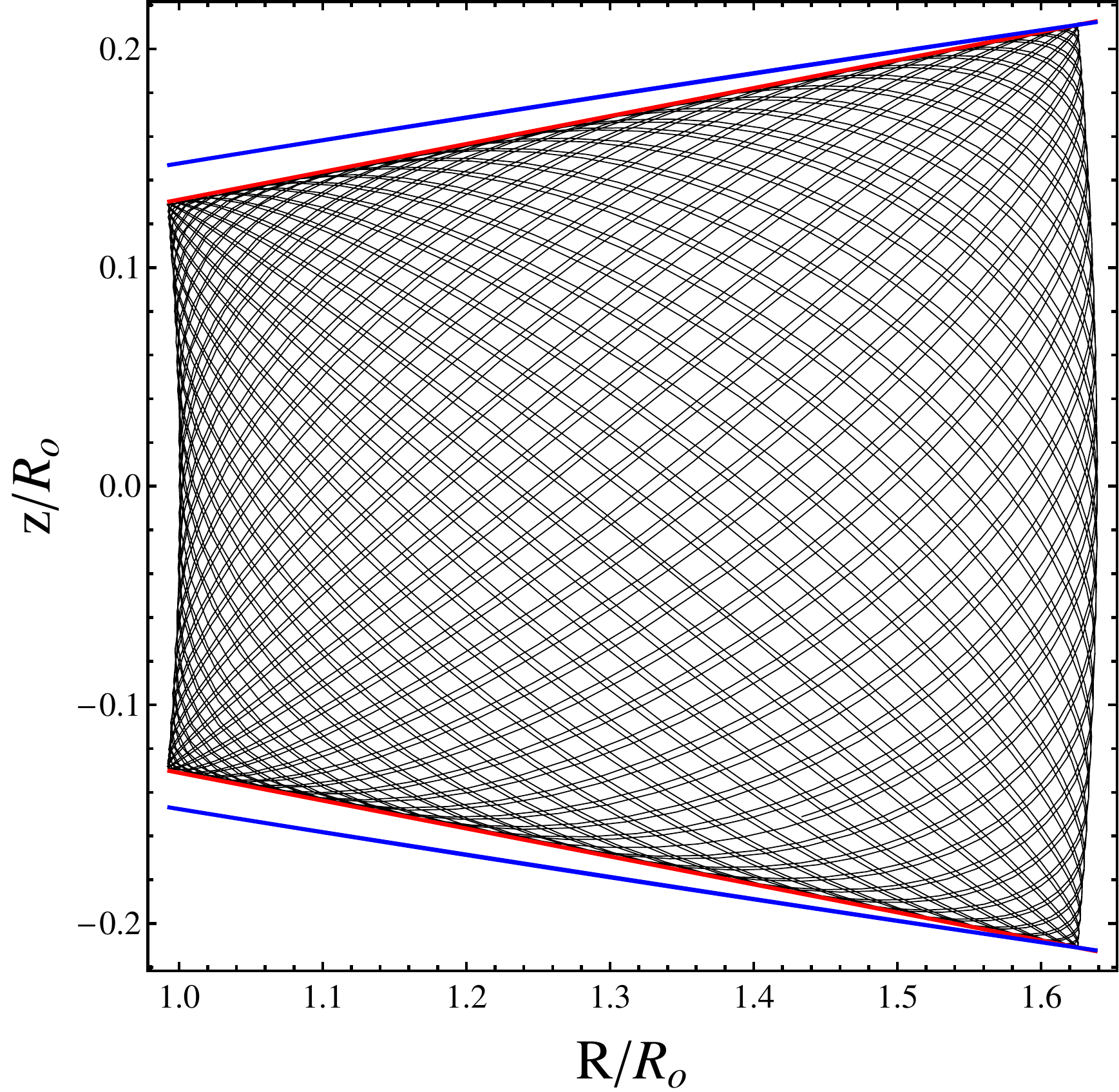}
\end{array}
$$
\caption{Envelopes for an orbit  around a monopole plus quadrupole central body  with $p=0.01$, $E/\alpha=-0.38$, $\ell/\sqrt{\alpha R_{o}}=1.1$, and
$a=b=10^{-8}$, corresponding to initial conditions $z(0)/R_{o}=10^{-10}$,
$V_R (0)=0$ and:  (a) $R(0)/R_{o}=1.4$, (b) $R(0)/R_{o}=1.2$, (c) $R(0)/R_{o}=1.6$, (d) $R(0)/R_{o}=1.0$.} \label{fig:2}
\end{figure*}

\begin{figure*}
$$
\begin{array}{cc}
  (a) & (b)\\
  \epsfig{width=6cm,file=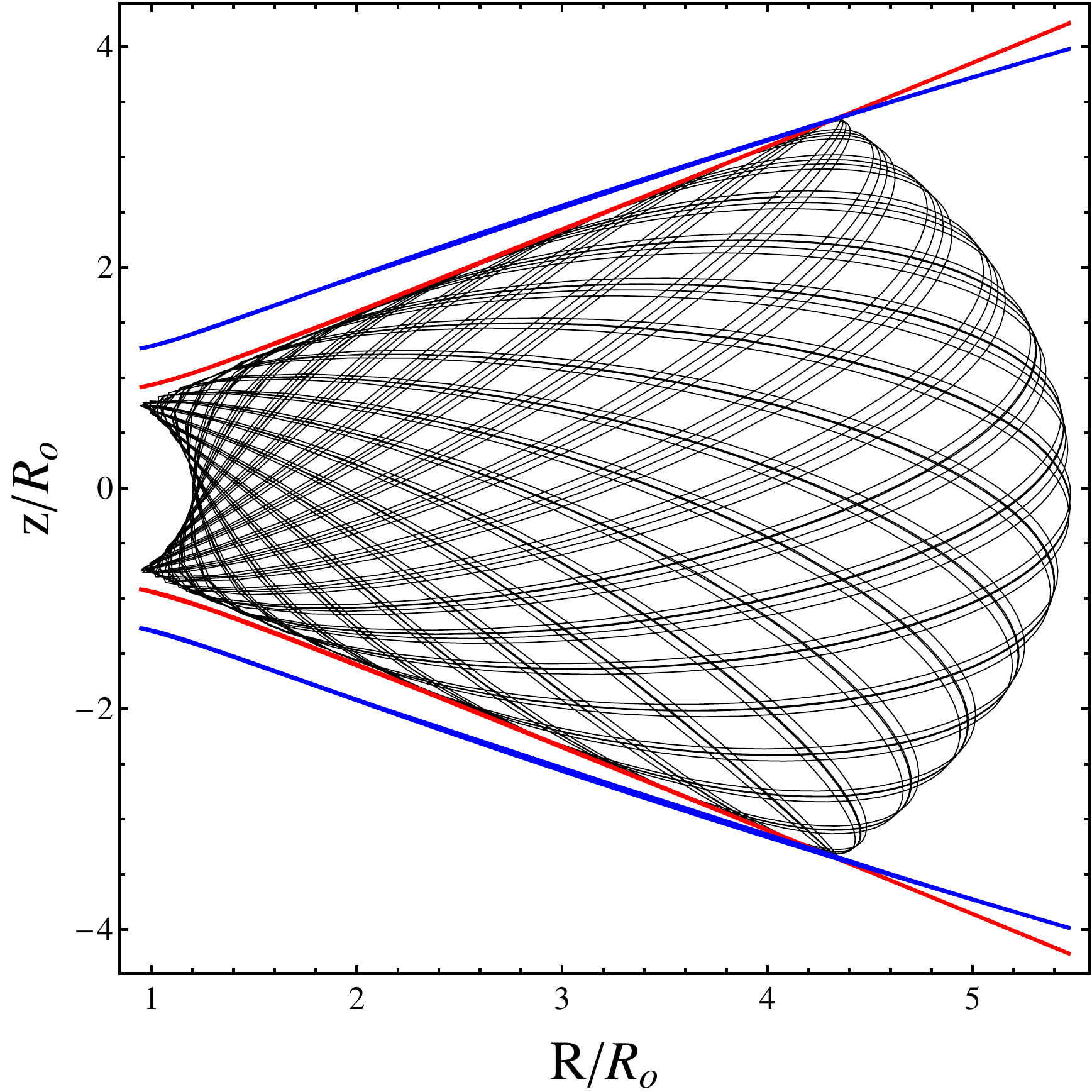} & \epsfig{width=6cm,file=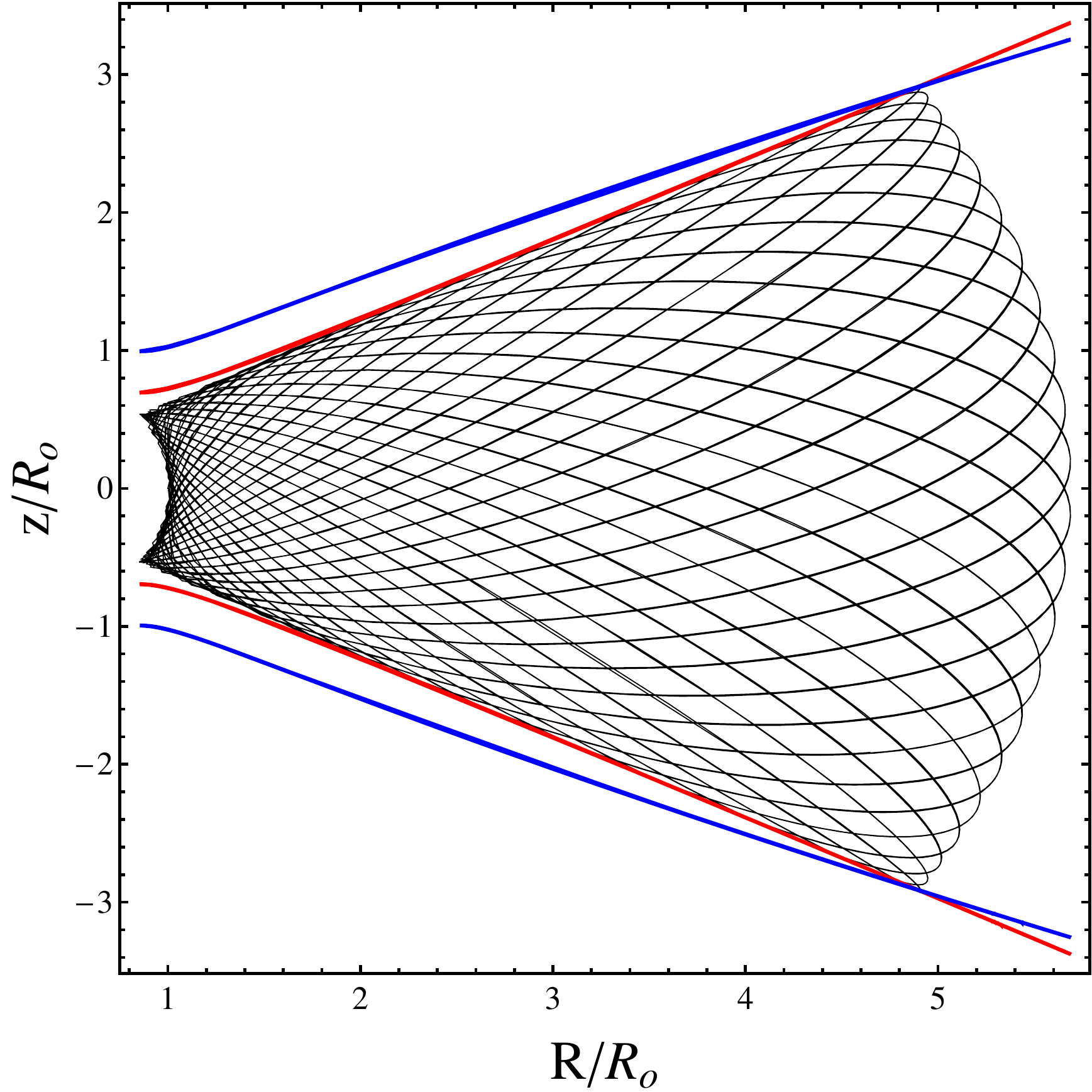}
\end{array}
$$
\caption{Same as figure~\ref{fig:2} but with $p=0.1$, $E/\alpha=-0.15$, $\ell/\sqrt{\alpha R_{o}}=1.1$, and
$a=b=10^{-8}$, corresponding to initial conditions $z(0)/R_{o}=10^{-10}$,
$V_R (0)=0$ and:  (a) $R(0)/R_{o}=1.2$, (b) $R(0)/R_{o}=1.0$.
} \label{fig:3}
\end{figure*}

We also note that (\ref{3}) does not work for (i) $a=0$, independently of the values for $b$; (ii) $b=0$, independently of the values for $a$.
But it works very well for $a\rightarrow 0$ and $b\rightarrow 0$ simultaneously, i.e. for values arbitrarily close to zero, without actually being zero. This
is illustrated in the orbits of figure \ref{fig:2}, whose vertical amplitudes (larger than the ones in figure \ref{fig:1}) are well reproduced after choosing $a=b=10^{-8}$. Indeed,
we observed that the predictions of (\ref{3}) were quite accurate even for orbits with vertical amplitudes up to 10 times
greater than those shown in figures~\ref{fig:1} and \ref{fig:2}, as the orbits shown in figure~\ref{fig:3}. In such a case, we only observe significant
deviations from formula (\ref{3}) for  $R/R_o \approx 1$, i.e. near the source.

Since the best approximations of the form (\ref{3}) to the numerically calculated envelopes are obtained for $a\rightarrow 0$ and $b\rightarrow 0$ simultaneously, we can simplify our procedure by using only
one parameter in expressions (\ref{Miyamoto-trans})--(\ref{3}), instead of the two $a$ and $b$.

\begin{figure*}
$$
\begin{array}{cc}
  (a) & (b)\\
  \epsfig{width=6cm,file=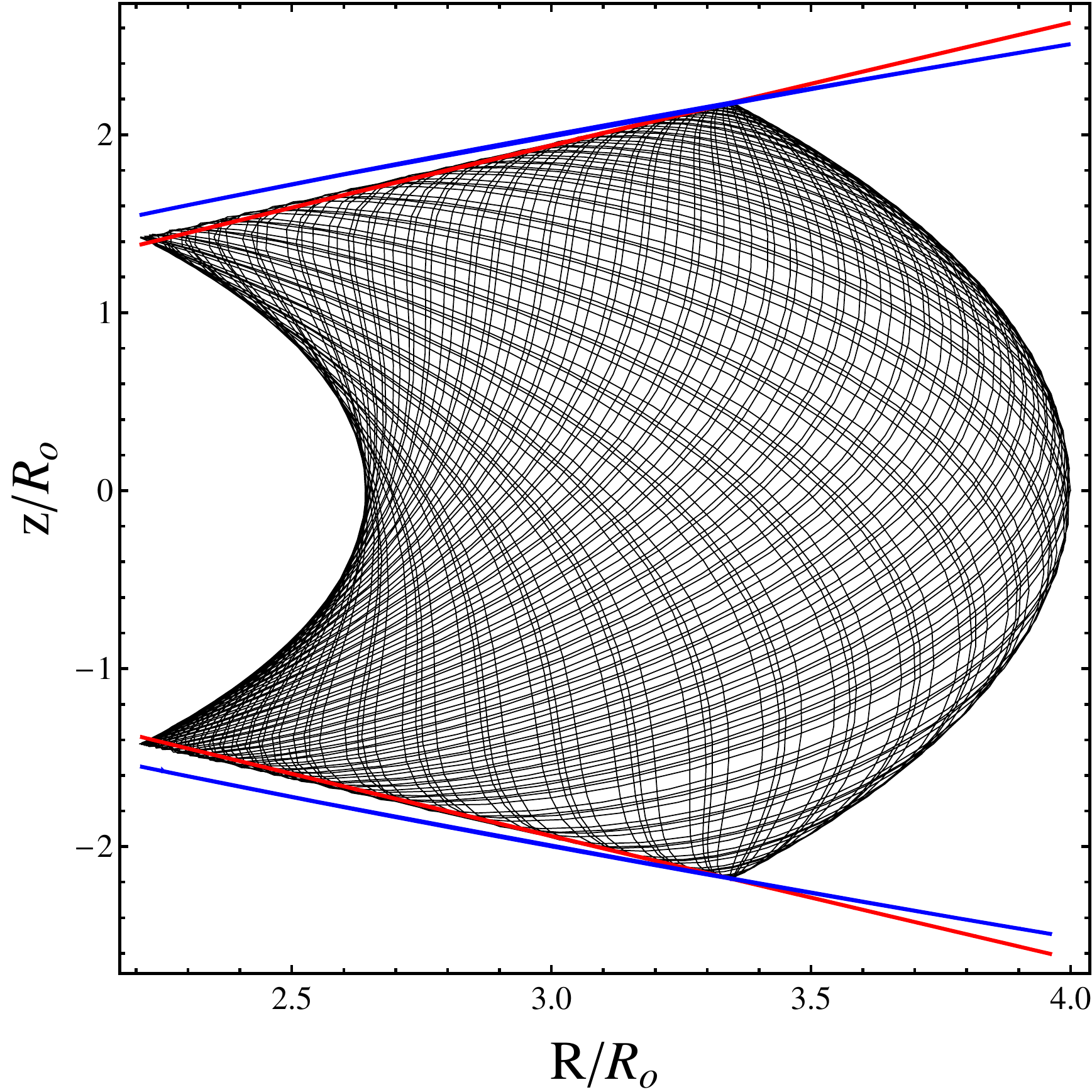} & \epsfig{width=6cm,file=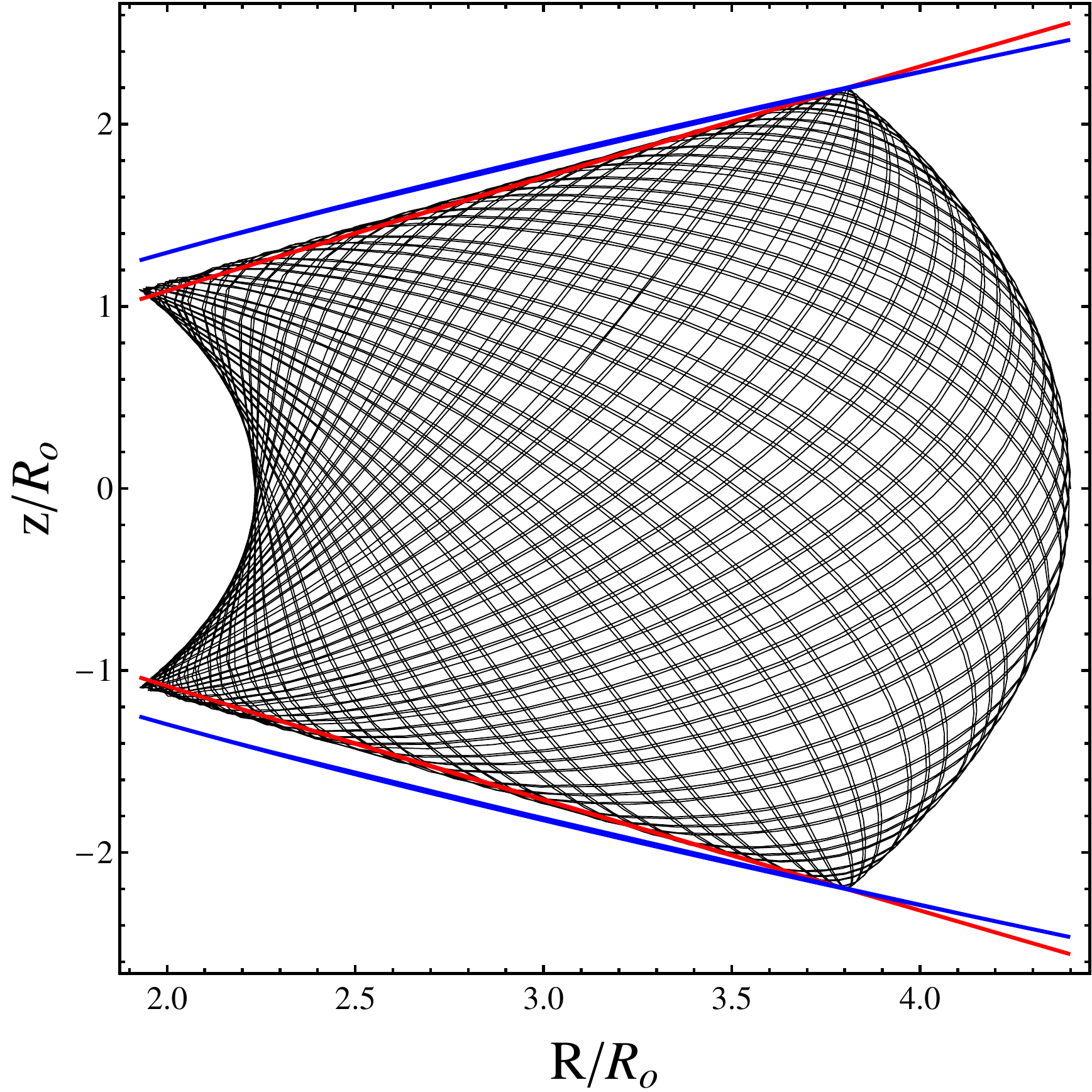}
\end{array}
$$
\caption{Envelopes for two orbits  around a Kalnajs disc of radius $R_o$ and choosing
 $\ell\sqrt{R_{o}^{3}/GM}=1.5$, $E R_o/GM=-0.15$,  $z(0)/R_{o}=10^{-10}$, $V_R (0)=0$, (a) $R(0)/R_{o}=4.0$ and (b) $R(0)/R_{o}=4.4$.
 The numerically integrated orbit in the meridional plane $Rz$ is presented in black; the
 blue curve is the prediction of adiabatic approximation (\ref{adiabatic})
 and the red curve represents the envelope predicted by (\ref{3})
 when we choose $a = b =  10^{-8}$ .} \label{fig:4}
\end{figure*}

Another situation where we can test the predictions of (\ref{3}) (by considering $a=b$) is the potential of a thin disc with finite extension, as the well-known Kalnajs disc (see for example \citealp{caosKalgen}):
\begin{equation}\label{Kalnajs}
   \Phi=-\frac{M G}{R_{o}}\left[\cot^{-1}\zeta + \frac{1}{4}\left((3\zeta^{2}+1)\cot^{-1}\zeta - 3\zeta\right)(3\eta^{2}-1)\right],
\end{equation}
where $M$ is the disc mass, $R_{o}$ is the disc radius and
\begin{equation}\label{oblatas1}
   \zeta=\frac{\mbox{Re}\left[\sqrt{R^{2}+(z-i R_{o})}\right]}{R_{o}},
\end{equation}
\begin{equation}\label{oblatas2}
   \eta=-\frac{\mbox{Im}\left[\sqrt{R^{2}+(z-i R_{o})}\right]}{R_{o}}.
\end{equation}
This case, as well as the  potential of any axisymmetric thin disc with finite extent,
takes into account all the terms in the multipolar expansion (\ref{2}).
 The results are similar to those found in the previous case, as exemplified in figure \ref{fig:4}, corresponding to two typical box orbits with vertical amplitude of the order of the disc radius. Again we perceive that the prediction  of (\ref{3}) significantly improves the prediction of adiabatic approximation.

\section{Conclusion and perspectives}\label{sec:conclusion}

Given an axially symmetric solution of Laplace equation,  $\Phi(R,z)$, such that $\Phi(R,z)=\Phi(R,-z)$,
we can establish that the $z$-amplitude of boxed orbits crossing the $z=0$ plane is given approximately by the formula
\begin{equation}\label{thirdIntegral}
   Z(R)\propto\Sigma^{-1/3}_{\rm ef}(R;a\approx 0),
\end{equation}
with $a$ close to zero but finite (the precision of the approximation gets better as $a$ diminishes), where $\Sigma_{\rm ef}(R;a)$ is the effective integrated density at radius $R$, defined by
\begin{equation}\label{DensIntEff2}
\Sigma_{\rm ef}(R;a)\equiv \int_{-a}^{a}\rho_{\rm ef}(R,z;a)dz,
\end{equation}
with $\rho_{\rm ef}$, given by
\begin{equation}\label{effective-density2}
    \rho_{\rm ef}(R,z;a)=\frac{1}{4\pi G}\nabla^{2}\Phi(R, a+\sqrt{a^{2}+z^{2}}),
\end{equation}
being the effective density obtained after perform the transformation
\begin{equation}\label{Miyamoto-trans2}
   z\rightarrow a+\sqrt{a^{2}+z^{2}},\qquad (a >0),
\end{equation}
on the vacuum solution $\Phi(R,z)$.

The result presented above can be considered, at the same time,  an application and an extension of the
ideas introduced in the references \citet{vieira2,vieira,vieira3},  devoted to study the integrability of disc-crossing orbits,
but here we focus on test-particle motion that takes place entirely in vacuum. A natural next step in this direction is
to introduce this formalism in the realm of general relativity (as an extension of the envelopes in razor-thin disc
spacetimes obtained in \citealp{vieira4}), in order to describe envelopes of geodesics associated with
vacuum solutions of Einstein's equations with axial symmetry.

Another interesting step, on the other hand, has to do with the fact that the above result could be an important complement in the
 the discussion carried out by \citet{letelier2}, about chaotic and regular motion in ``monopole $+$ oblate quadrupole'' potentials.
It would be interesting to find the explicit form for the third integral of motion associated with equation (\ref{3}) in terms of the phase-space coordinates
$(R,z, P_R ,  P_z)$, in order to provide predictions for the corresponding Poincar\'{e} surfaces of section (among other dynamical quantities), which  would have applications in both astrophysics and nuclear physics.


\section*{Data availability}

The data underlying this article will be shared on reasonable request to the corresponding author.

\end{document}